\documentclass[11pt]{article} 
\usepackage{amsfonts,amssymb,slashed,latexsym,setspace}
\usepackage{graphicx,graphics,amssymb,epsf,rotate,subfigure} 
\usepackage{times,cite,color,epsfig,epsf}
\usepackage{multirow}
\usepackage{hyperref}
\usepackage{bm}
\begin{document}

\begin{flushright}
SINP/TNP/2013/07
\end{flushright}

\begin{center}
{\Large \bf Minimal supersymmetry confronts $\bm{R_b}$, $\bm{A^b_{\rm
      FB}}$ and $\bm{m_h}$} \\
\vspace*{1cm} \renewcommand{\thefootnote}{\fnsymbol{footnote}} { {\sf
    Gautam Bhattacharyya${}^{1}$}, {\sf Anirban Kundu${}^{2}$} and
  {\sf Tirtha Sankar Ray${}^{3}$}} \\
\vspace{10pt} {\small ${}^{1)}$ {\em Saha Institute of Nuclear
    Physics, 1/AF Bidhan Nagar, Kolkata 700064, India} \\ 
${}^{2)}$ {\em Department of Physics, University of Calcutta, 92
    Acharya Prafulla Chandra Road, Kolkata 700009, India} \\ 
${}^{3)}$ {\em ARC Centre of Excellence for Particle Physics at the Terascale,
School of Physics, \\ University of Melbourne, 
Victoria 3010, Australia}} 
\normalsize
\end{center}

\begin{abstract}
We study the impact of the measurements of three sets of observables
on the parameter space of the constrained minimal supersymmetric
Standard Model (cMSSM), its slightly general variant, the
non-universal scalar model (NUSM), and some selected benchmark points
of the 19-parameter phenomenological MSSM (pMSSM): ($i$) the direct
measurement of the Higgs boson mass $m_h \approx 125$ GeV at the CERN
Large Hadron Collider (LHC); ($ii$) $Z$ boson decay width in the $b
\bar{b}$ channel normalized to its hadronic width ($R_b$), and the
forward-backward asymmetry on the $Z$-peak in the same channel
$(A_{\rm FB}^b)$; and ($iii$) several $B$-physics observables, along
with $(g-2)$ of muon. In addition, there are constraints from
non-observation of superparticles from direct searches at the LHC.  In
view of the recently re-estimated standard model (SM) value of $R_b$
with improved higher order corrections, the measured value of $R_b$
has a 1.2$\sigma$ discrepancy with its SM value, while the
corresponding discrepancy in $A_{\rm FB}^b$ is 2.5$\sigma$.  MSSM
contributions from light superpartners improve the agreement of $R_b$
but worsen that of $A_{\rm FB}^b$. We project this ($R_b$-$A_{\rm
  FB}^b$) tension {\em vis-\`{a}-vis} the constraints arising from
other observables in the parameter space of cMSSM and NUSM. We also
consider a few well-motivated pMSSM benchmark points and show that pMSSM
does not fare any better than the SM.

\end{abstract}

\setcounter{footnote}{0}
\renewcommand{\thefootnote}{\arabic{footnote}}

\section{Introduction}
It is still premature to conclude that the recently discovered scalar
particle with a mass of around 125 GeV at the LHC
\cite{cms-higgs,atlas-higgs} is the standard model (SM) Higgs
boson. Among other possibilities, it can very well be the lightest
neutral Higgs boson of the minimal supersymmetric standard model
(MSSM), which is the most advocated beyond the SM scenario. Be that as
it may, stringent constraints apply on the supersymmetric parameter
space, which are, at least for the minimal versions, more severe than
those placed from the non-observation of superparticles from direct
searches at the colliders. In spite of several advantages that
supersymmetry offers, like the solution of the big hierarchy problem
and even the provision of a cold dark matter candidate, we cannot
escape the following pertinent question: does supersymmetry even
partially ease the existing tension between the SM predictions and the
experimental measurements for some specific observables? The tension
with $(g-2)$ of muon, namely, $\delta a_\mu = a_\mu^{\rm exp} -
a_\mu^{\rm SM} = (28.7 \pm 8.2) \times 10^{-10}$, i.e. a 3.5$\sigma$
deviation of the measured value \cite{muon_g_exp} from its SM
expectation \cite{g-2_hagiwara2011,g-2_davier2010}, is an old
intriguing one. Similarly, there is a longstanding $2.5\sigma$
discrepancy between the SM expectation and the experimental value for
the $e^+e^-\to b\bar{b}$ forward-backward asymmetry $A^b_{\rm FB}$
measured on the $Z$-peak \cite{pdg}. Recently, a 1.2$\sigma$
discrepancy, but with an opposite pull to that of $A^b_{\rm FB}$,
between the SM prediction and the experimental value for $R_b = \Gamma
(Z \to b\bar b) / \Gamma (Z \to {\rm hadrons})$ has been reported
following the latest SM calculation taking electroweak two-loop plus
order $\alpha \alpha_S^2$ three-loop effects \cite{rb-theory}.
Although 1.2$\sigma$ is not statistically too much significant, it is
this opposite pull between $R_b$ and $A^b_{\rm FB}$ that we would like
to exploit for discriminating supersymmetric models.  Possible ways to
ease these hiccups in a model-independent effective theory approach
and the corresponding LHC signatures have recently been discussed
\cite{Choudhury:2013jta}. But, what these tensions imply for
supersymmetric models is the main concern of the present paper.

For definiteness, we will start with the parameter space of the
constrained MSSM, called cMSSM, completely specified at the grand
unification (GUT) scale by a common scalar mass ($m_0$), a common
gaugino mass $(M_{1/2})$, a universal scalar trilinear parameter
($A_0$), the ratio of the vacuum expectation values of the two Higgs
doublets ($\tan\beta$), and the sign of the Higgsino mass parameter
($\mu$) \cite{susy-books,Martin:1997ns}. We place indirect constraints
on the cMSSM parameter space from observables classified here under
three categories:
\begin{enumerate} 
 \item The lightest CP-even Higgs mass $m_h \approx 125$ GeV, reaching
   out to which requires the stop squarks to be very heavy (several
   TeV) or having a substantial mixing between their left and right
   components \cite{OYY,Okada:1990gg}.  Since all superparticle masses
   are interrelated in terms of the $(4+1)$ GUT scale parameters, the
   Higgs mass puts by far the strongest constraint on the cMSSM
   parameter space.

\item A set of $B$-physics observables, namely, ${\rm Br}(b\to s
  \gamma)$, ${\rm Br}(B_s \to \mu^+\mu^-)$, ${\rm Br}(B^\pm \to
  \tau\nu)$, and ${\rm Br} (B \to D^{(*)} \tau\nu)$. The first two
  processes, being loop-induced in both SM and cMSSM, put a strong
  constraint on the cMSSM parameters from the fact that the data on
  them are quite precise and completely consistent with the SM
  expectations. The new physics parts for the remaining ones are
  mediated essentially by the charged Higgs boson, and here the data
  show some intriguing discrepancy with the SM expectations.  We
  include the measurement of $(g-2)$ of muon too in this category,
  which prefers the smuons and gaugino/higgsino to be light.

\item
The $Z$-peak observables $R_b$ and $A_{\rm FB}^b$. Both are precision
electroweak observables, with a pull of 1.2$\sigma$ and 2.5$\sigma$
respectively, when experimental data are compared with their SM
predictions. The supersymmetric contributions to these observables in
the current context form the core of our analysis. 
\end{enumerate}

Constraints from the interplay of the first two categories listed
above have already been placed by several authors
\cite{Strege:2012bt,cmssm-others}.  The main thrust of the present
paper is to explore if it is possible to additionally satisfy the
constraints listed in the third category. The electroweak precision
data was analyzed vis-\`a-vis MSSM in Ref.~\cite{Heinemeyer:2007bw},
but at a time when the discrepancy between the measured value of $R_b$
and the SM one-loop estimate was only about $0.6\sigma$, and the Higgs
boson was not discovered. So it is only proper to readdress the issue
in the light of the present wisdom, especially when the Higgs boson
has finally been observed, and estimate of higher-order effects on
$R_b$ has widened the discrepancy from $0.6\sigma$ to
$1.2\sigma$. Also, the opposite pull between $R_b$ and $A_{\rm FB}^b$
should be paid due attention, as these are precision observables.

We note that in cMSSM the masses of the squarks and the charged Higgs
boson stem from a common $m_0$.  Therefore, non-observation of squarks
in direct searches automatically implies that the charged Higgs boson
has to be heavy.  For this reason, for some of the $B$-physics
observables where the charged Higgs boson contributes as the sole new
physics at tree level, its numerical impact is rather suppressed.  To
liberate the charged Higgs boson from this unfavorable tie-up with the
squarks, we consider a slightly wider variant, called the
non-universal scalar model (NUSM), which is characterized by two
different soft scalar masses at the GUT scale: $m_{16}$ for the matter
scalars which belong to a 16-plet, and $m_{10}$ for the Higgs scalars
which belong to a decuplet, of the underlying SO(10)
\cite{nusm-1}. Note that even if there is a common scalar mass at the
Planck scale $M_{Pl}$, the matter and Higgs scalar masses at the GUT
scale $M_G$ can be different because of renormalization group running
controlled by {\em a priori} unknown physics between $M_{Pl}$ and
$M_G$. From this perspective, one might take $m_{10}$ and $m_{16}$ as
independent parameters.  In this case, non-observation of squarks does
not necessarily imply that the charged Higgs boson is
heavy\footnote{This model is not the same as the so-called
  non-universal Higgs models (NUHM) \cite{NUHM-conven} which contain
  two additional free parameters, namely, $\mu$ and $m_A$. For NUSM,
  the weak scale parameters $m_{H_u}^2$ and $m_{H_d}^2$ are not
  independent as at GUT scale these two parameters are the same as
  $m_{10}^2$.}. Still, the masses of the charged Higgs boson and the
squarks in NUSM are not fully uncorrelated because of the radiative
corrections to the Higgs mass parameters coming from the stop squarks
involving large Yukawa couplings.  Even then, we observe that a lot of
parameter space in the charged Higgs mass plane which is disfavored in
cMSSM is resurrected in NUSM.

In the last part of this paper, we will consider selected benchmark
points of the 19-parameter phenomenological MSSM (pMSSM)
\cite{pmssm-ref}, which is a subspace of pMSSM with parameters defined
at the weak scale, without introducing any theoretical prejudice about
their high scale behaviors. 24 such benchmark model points have been
considered in Ref.\ \cite{cahill}, which correspond to different
regions of the MSSM parameter space, consistent with all other
constraints like the Higgs mass, dark matter relic density, and direct
searches at the LHC. Out of these 24 model points, 5 have been
shortlisted in \cite{cahill}. We will analyse how these well-motivated
benchmark points react to the combined effects of ($R_b$--$A_{\rm
  FB}^b$) and a few other observables.

\section{${\mathbf{R_b}}$ and ${\mathbf{A_{\rm FB}^b}}$}
The one-loop corrected $Zb\bar{b}$ coupling, including new physics,
can be written as
\begin{equation}
{\cal L}_{Zb\bar{b}} = - \frac{g}{\cos\theta_W}
\bar{b}\gamma^\mu\left[(g_L^b+\delta g_L^b)P_L + (g_R^b + \delta
  g_R^b)P_R\right]b Z_\mu \, ,
\end{equation}
where 
\begin{equation}
g_L^b = \sqrt{\rho_b}\left(t_3^b - \kappa_b Q_b \sin^2\theta_W \right)\,,\ \
g_R^b = - \sqrt{\rho_b} \kappa_b Q_b \sin^2\theta_W\,,
\end{equation}
with $\rho_b = 0.9869$ and $\kappa_b = 1.0067$ after incorporating the
SM electroweak corrections in the $\overline{\rm MS}$ scheme
\cite{pdg}, and $\delta g_{L,R}^b$ capture the higher-order effects
coming from new physics.  Here, $t_3^b=-\frac12$, $Q_b = -\frac13$,
and $P_{L(R)} = \frac12(1-(+)\gamma_5)$.  The loop corrections induced
by the superparticles can be approximated as
\begin{eqnarray}
\delta g_{L(R)}^b &=& \xi F_{L(R)}\,, ~~~{\rm with}~~
\xi = \frac{\alpha}{4\pi\sin^2\theta_W}\, .
\end{eqnarray}
The expressions for the quantities $F_L$ and $F_R$ have been adapted
from Appendix B of Ref.~\cite{Boulware:1991vp}. They contain
contributions from top-charged Higgs, stop-chargino and
sbottom-neutralino loops, expressed in terms of the Passarino-Veltman
functions \cite{Passarino:1978jh}. The new contribution to $R_b$ can
be written as
\begin{equation}
\delta R_b = R_b^{\rm SM}\left(1-R_b^{\rm SM}\right) \nabla_b \, , 
\end{equation}
with
\begin{equation}
\nabla_b = \xi\left[ \frac{2g^b_L F_L + 2 g^b_R
    F_R}{\left(g^b_L\right)^2 + \left(g^b_R\right)^2}\right] \approx
\xi \left[-\frac{60}{13} F_L + \frac{12}{13} F_R\right] \, , 
\label{rb-susy}
\end{equation}
where the right-most simplified expression in Eq.~(\ref{rb-susy}) is
shown only for illustration by approximating $\kappa_b = 1$ and
$\sin^2\theta_W = 0.25$. {\em However, this approximation is shown
  only for illustrative purpose}. The full expression in MSSM can be
found in \cite{Boulware:1991vp,Heinemeyer:2007bw} which we have
employed for our numerical analysis..  The Gfitter group
\cite{gfitter} has recently updated the SM fit (after the Higgs
discovery) using the improved calculation of $R_b$ with the result
that the discrepancy with the measured value is now 1.2$\sigma$
\cite{rb-theory}. The new situation is the following:
\begin{equation}
R_b~({\rm exp}) = 0.21629\pm 0.00066\,,\ \ 
R_b~({\rm SM}) = 0.21550\pm 0.00003 \, .
\label{rb-data}
\end{equation}

The change in $A^b_{\rm FB}$ can similarly be written as 
\begin{equation}
\frac{A^b_{\rm FB}}{A^b_{\rm FB}({\rm SM})} - 1 = \xi \left[\frac{2g^b_R F_R -
    2g^b_L F_L} {\left(g^b_R\right)^2 - \left(g^b_L\right)^2} -
  \frac{2g^b_R F_R + 2g^b_L F_L} {\left(g^b_R\right)^2 +
    \left(g^b_L\right)^2}\right] \approx -\frac{5}{13}\xi \left[F_L +
  5 F_R\right]\, ,
\label{afb-susy}
\end{equation}
for which a $2.5\sigma$ discrepancy between the SM and experimental
values is known to exist for a while \cite{pdg}:
\begin{equation}
A^b_{\rm FB}~({\rm exp}) = 0.0992\pm 0.0016\,,\ \ {\rm and} \ \ 
A^b_{\rm FB}~({\rm SM}) = 0.1032^{+0.0004}_{-0.0006}\,.
\label{afb-data}
\end{equation}

A comparative study of $R_b$ and $A^b_{\rm FB}$ is now in order.  It
is not difficult to understand the tension between their
supersymmetric contributions irrespective of any other constraints.
In MSSM (i.e. not just in cMSSM), both $F_L$ and $F_R$ turn out to be
negative, with $|F_L| > |F_R|$, because of the presence of the
numerically dominant stop-chargino loop contribution for $F_L$, which
is absent for $F_R$.  Thus, one can have a consistent solution for
$R_b$, as can be seen from Eqs.~(\ref{rb-susy}) and (\ref{rb-data}).
On the other hand, as Eq.~(\ref{afb-susy}) shows, the supersymmetric
contribution to $A^b_{\rm FB}$ is always positive, while according to
Eq.~(\ref{afb-data}) experimental data prefers a negative
contribution. Thus, not only cMSSM but a large class of supersymmetric
model for which $F_L, F_R < 0$ and $|F_L| > |F_R|$ cannot
simultaneously satisfy $R_b$ and $A^b_{\rm FB}$.

To numerically appreciate the combined effects of $R_b$ and $A^b_{\rm
  FB}$, we first define what is known as `pull' for a given observable
($O$), as
\begin{equation}
{\rm Pull} = \frac{O_{\rm theory} - O_{\rm expt}}{\sigma_{\rm expt}} \,,
\label{eq:def-pull}
\end{equation}
where $O_{\rm expt}$ is the experimental central value and
$\sigma_{\rm expt}$ is one standard deviation experimental error.
Then we introduce a quantity $X_b$ defined as 
\begin{equation}
X_b =  \sqrt{({\rm Pull~of}~R_b)^2 + ({\rm Pull~of}~A^b_{\rm FB})^2} \, .
\label{eq:Xb}
\end{equation}
Eqs.~(\ref{rb-data}) and (\ref{afb-data}) tell us that the pulls of
$R_b$ and $A_{\rm FB}^b$ are $-1.2$ and $2.5$, respectively, so that
$X_b^{\rm SM} = 2.8$.  Light superparticles can improve the agreement
in $R_b$, but any such contributions can only worsen the tension in
$A^b_{\rm FB}$, as explained earlier. Thus $X_b$ cannot be smaller
than $2.5$.  This motivates us to look for the parameter space where
$X_b$ lies between 2.5 and its SM value 2.8 (which corresponds to
completely decoupled supersymmetry). For illustration, we have chosen
$X_b = 2.6$ and 2.7 as two representative values, in the sense that
for each case we accept the parameter space which yields $X_b$ less
than that value. We recall in this context that although $R_b$ and
$A_{\rm FB}^b$ both depend on the dynamics of the $Zb\bar{b}$ vertex,
the former is a ratio of decay widths and the latter is an asymmetry,
and they are experimentally independent observables. Hence, $X_b$ is a
good measure of the combined tension produced by these two
observables.

\section{Results}
\subsection{cMSSM and NUSM}
The model parameters have been scanned in the following range: $0 <
\left(m_0, M_{1/2}\right) < 2$, $-7 < A_0 < 0$ (all in TeV), $10 \leq
\tan\beta \leq 45$. For the NUSM scenario, we scan over the additional
parameter $m_{10}$ in the range of [0:2] TeV.  We have used
SuSpect2.41 \cite{Djouadi:2002ze} to generate the weak scale spectrum
from the high scale boundary values. The package micrOMEGAs2.4
\cite{Belanger:2010gh} has been utilized to extract observables like
branching ratios of $b \to s \gamma$, $B_{s} \rightarrow \mu^+ \mu^-$
and $B^+\rightarrow \tau^+ \nu_{\tau}$.  We have used the following
experimental ranges for the $B$-physics observables
\cite{Amhis:2012bh,Aaij:2012nna}, admitting $2\sigma$ spread around
their central values, except for ${\rm Br}(B_d \to \mu^+ \mu^-)$ which
is taken at $90\%$ confidence limit:
\begin{eqnarray}
 {\rm Br}(b \to s \gamma) &=& (3.55 \pm 0.24 \pm 0.09)\times
 10^{-4}\,, ~~~ {\rm Br}(B_s \to \mu^+ \mu^-) =
 \left(3.2^{+1.5}_{-1.2}\right) \times 10^{-9}\,, \nonumber\\ {\rm
   Br}(B_d \to \mu^+ \mu^-) &<& 0.94 \times 10^{-9}\,, ~~~{\rm Br}(B^+
 \to \tau^+ \nu) = (1.15 \pm 0.23) \times 10^{-4} \, .
\end{eqnarray}

\begin{figure}
       \includegraphics[width=8.0cm]{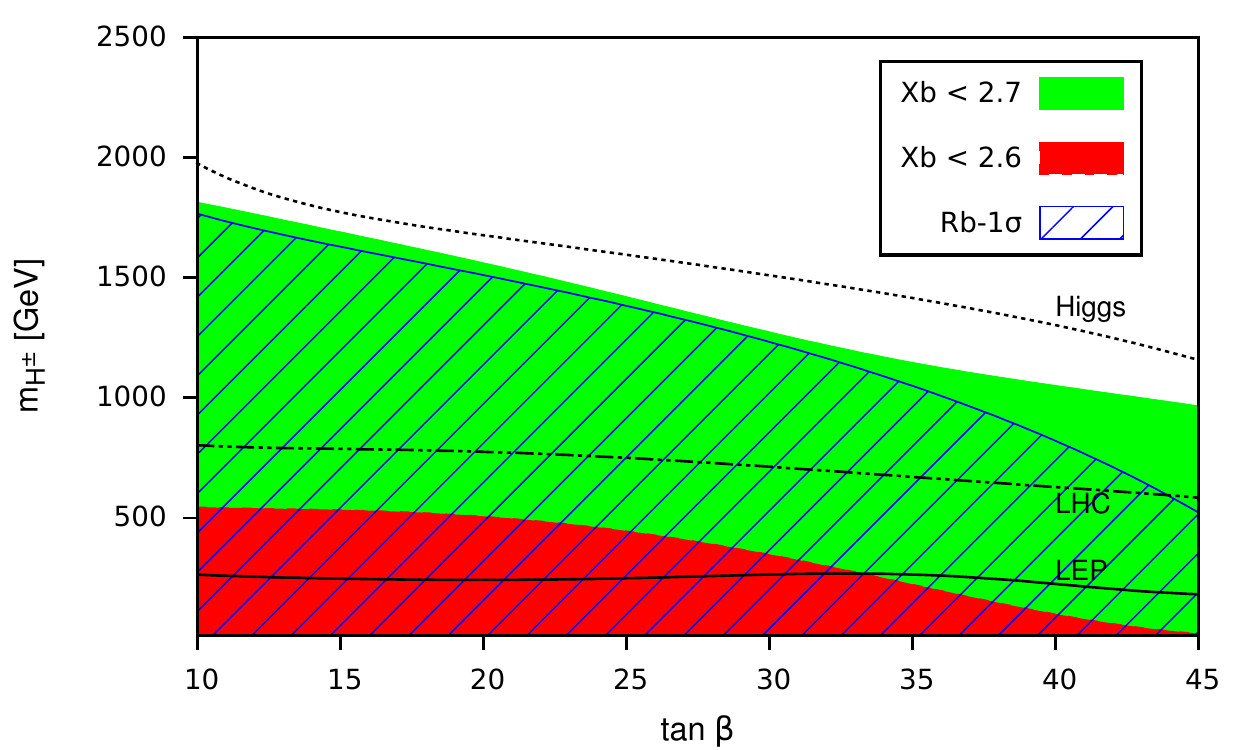}~~
       \includegraphics[width=8.0cm]{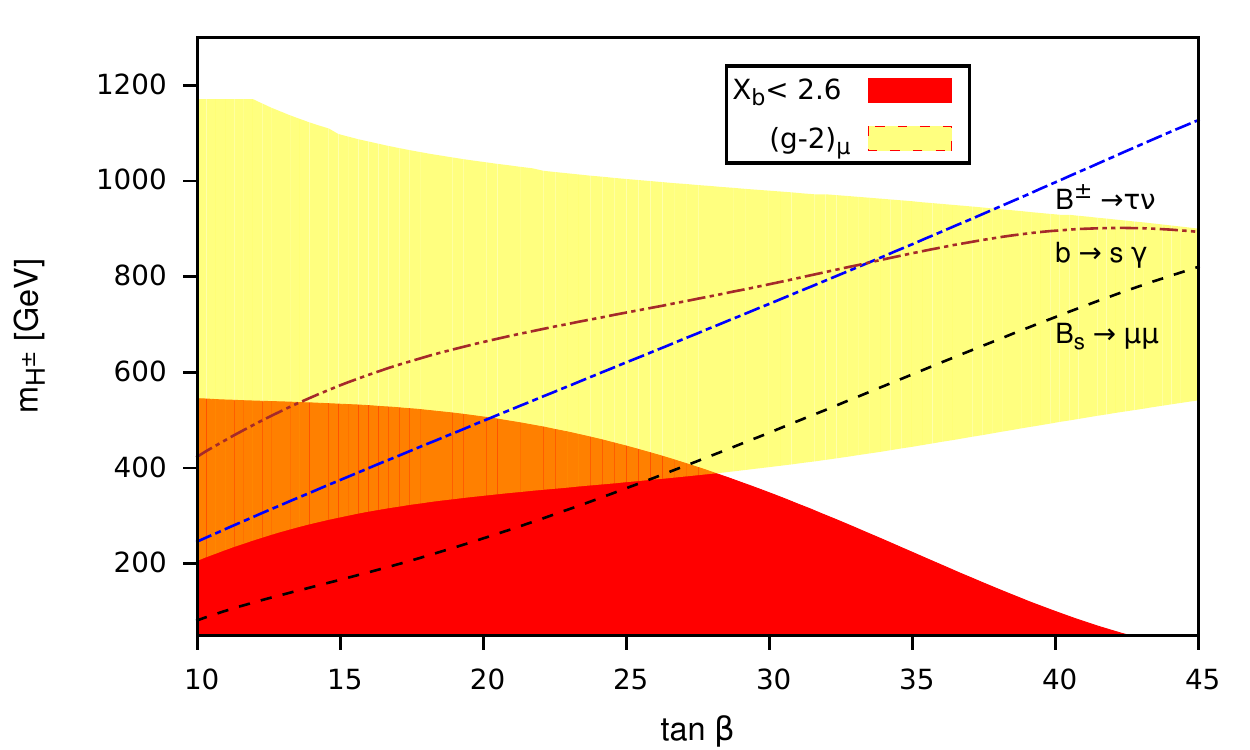}
\caption{\small (a) Left panel: Constraints on the
  $m_{H^+}$-$\tan\beta$ plane from $m_h = 123$ GeV, direct searches of
  gluinos and squarks at the LHC, and the LEP limit on the chargino
  mass.  The region above each line is allowed. The regions which
  satisfy $R_b$ at $1\sigma$ (blue dashed lines), $X_b < 2.6$
  (red/dark grey patch) and $X_b < 2.7$ (green/light grey patch) are
  separately shown. (b) Right panel: Allowed regions from several
  $B$-physics observables, $(g-2)$ of muon, all taken at
  $2\sigma$. The red (dark grey) and orange (medium grey) regions
  satisfy $X_b < 2.6$, while the yellow (light grey) and orange
  (medium grey) regions are allowed by $(g-2)$ of muon. For the
  $B$-physics observables, the region above each line is allowed.}
\label{fig:cmssm}
 \end{figure}

Our results for cMSSM are shown in the left and right panels of
Fig.~\ref{fig:cmssm} in the plane of the charged Higgs mass and
$\tan\beta$.  In both panels, we indicate the region which satisfy our
criteria for $X_b$. {\em In the left panel (a)}, besides showing the
effect of satisfying the Higgs mass, we display the constraints coming
from the non-observation of superparticles (mainly, the gluino and the
first two generations of squarks) from direct searches at the LHC
\cite{susy_sq_glu}, as well as those originating from the LEP limit on
the chargino mass \cite{pdg}.  The region which agrees with $R_b$
within 1$\sigma$ is shown by blue dashed lines.  While one might argue
that a discrepancy of $1.2\sigma$ is not of much significance, we
observe that softening the same to even $1\sigma$ requires such light
sparticles, in particular the stop squarks, as to be untenable with
the Higgs mass constraint.  We also show regions which admit $X_b <
2.6$ (red or dark grey) and $X_b < 2.7$ (green or light grey). We
reiterate that the 2.5$\sigma$ pull in $A_{\rm FB}^b$ essentially
controls the constraint from $X_b$.  {\em In the right panel (b)}, we
focus on the zone where constraints from several $B$-physics
observables, as well as the region allowed by $(g-2)$ of muon, are
prominent.  Wherever a single line is drawn for a particular
observable, the space above that line is allowed from the
corresponding constraint.  We make a cautionary remark that these
lines separate allowed and disallowed regions where all cMSSM input
parameters are marginalized.  Admittedly, we have not tried to
simultaneously fit all the electroweak observables, however, with such
large values of the sparticle masses, it is only natural to expect
that the observables already consistent with the SM would not suffer
any untenable tension.  The zone allowed by $(g-2)$ of muon is shown
by yellow patch.  The dominant contribution to $(g-2)$ of muon comes
when the bino and smuons floating inside the loops are light
\cite{Moroi:1995yh}.  We have kept the lightest Higgs boson mass at a
conservative value of 123 GeV, allowing for the possibility that
higher order radiative corrections may contribute a further 2 to 3 GeV
(setting $m_h=125$ GeV would result in even tighter constraints). Even
then, this rules out more region in the $m_H^+$ versus $\tan\beta$
plane than the other constraints. The reason is that one requires stop
masses as heavy as several TeV and a substantial mixing in the stop
mass-squared matrix to yield such a Higgs mass
\cite{Brummer:2012ns}. Among the flavor observables, $b \to s \gamma$
usually provides the strongest constraint \cite{Bertolini:1990if},
because it directly constrains the chargino mass which sits in the
lower end of the cMSSM spectrum. The constraint gets more stringent at
large $\tan\beta$ because the dominant chargino-stop loop amplitude
grows as $1/\cos^2\beta$. The supersymmetric contribution to the
branching ratio of $B_s \to \mu^+ \mu^-$ grows as $(\tan\beta)^6$
\cite{Arnowitt:2002cq}, and therefore at large $\tan\beta$ the
constraint is more stringent. As regards the process $B^+ \to \tau^+
\nu$, the dominant new physics contribution comes from tree-level
charged Higgs mediation, and for very large $\tan\beta$ provides a
stronger constraint than ${\rm Br}~ (b \to s \gamma)$. The constraint
essentially applies on the ratio $\tan\beta/m_{H^+}$, which is a
general feature of any two-Higgs doublet model \cite{Akeroyd:2003zr}.

\begin{figure}
       \includegraphics[width=8.0cm]{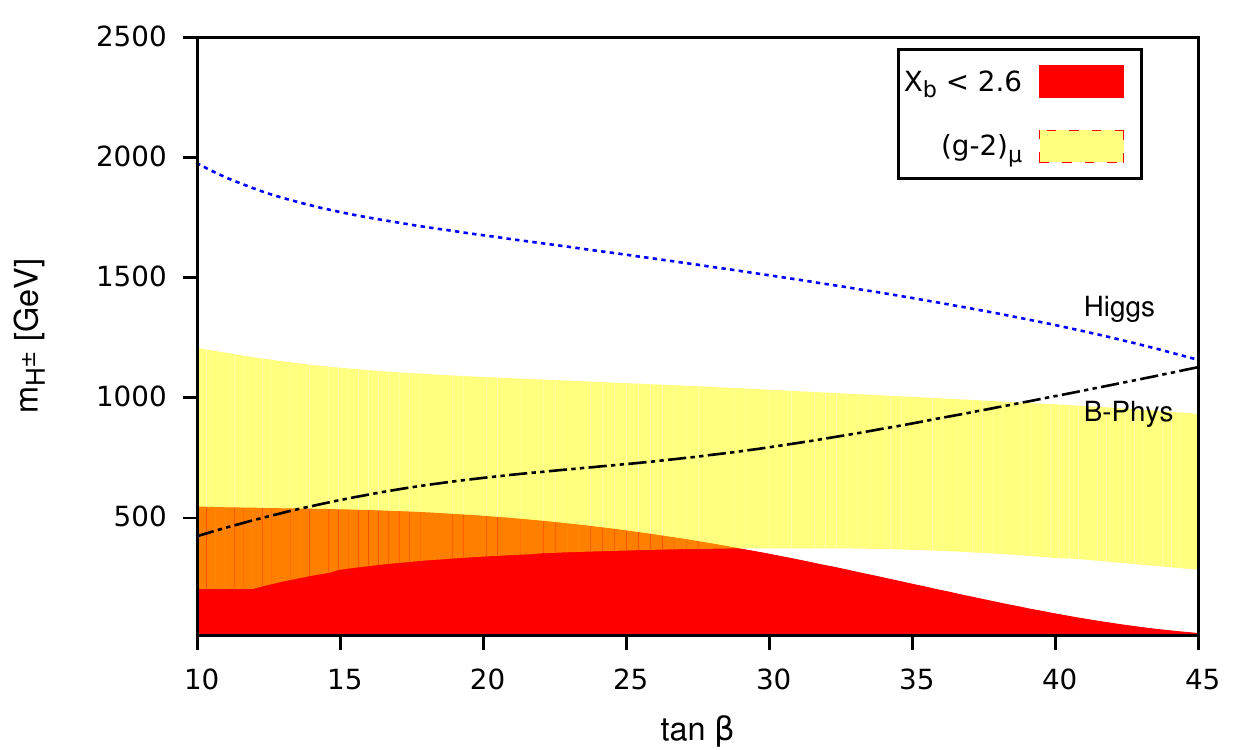}~~
       \includegraphics[width=8.0cm]{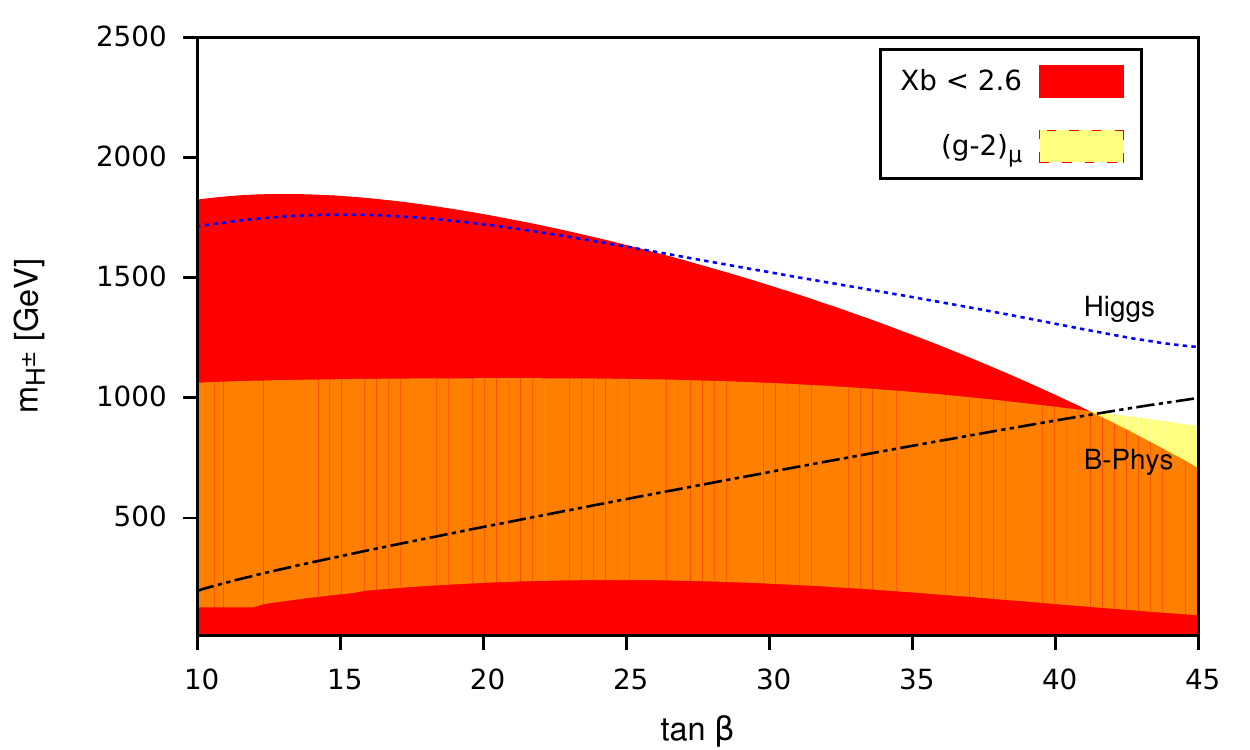}
\caption{\small A comparison of constraints between cMSSM (left panel)
  and NUSM (right panel) arising from the Higgs mass, $(g-2)$ of muon,
  the combined $B$-physics data, and the requirement of satisfying
  $X_b < 2.6$. The color codes are the same as in
  Fig.\ \ref{fig:cmssm}.}
\label{fig:compare}
 \end{figure}

In Fig.~\ref{fig:compare}, we compare cMSSM (left panel) with NUSM
(right panel). In each panel we show the impact of the combined
constraints coming from the Higgs mass, $B$-physics observables,
$(g-2)$ of muon, and $X_b$.  Constraints from the Higgs mass are
similar in the two cases in spite of the fact that in NUSM the charged
Higgs mass has its origin in $m_{10}$ while the squark masses are
unified at a different value $m_{16}$.  The reason can be traced to
the weak scale sum rule that holds in both cases, namely, $m^2_{H^+} =
M_W^2 + m_A^2$, where $m_A$ is the CP-odd Higgs mass. In both cases,
the satisfaction of $m_h \approx 123$ GeV requires $m_A$ to be very
large (the `decoupling' limit). A crucial observation is that unlike
in cMSSM there is a small region in NUSM which satisfies both $X_b <
2.6$ and the Higgs mass constraint.

We now comment on how the requirement of $X_b$ being less than some
representative value is transmitted to an upper limit on the charged
Higgs mass. This upper limit comes from the fact that the masses of
the charged Higgs and the stop squarks, both floating in independent
triangle loops of the effective $Zb\bar{b}$ vertex, are intimately
related through $m_0$ and $M_{1/2}$. For moderate $\tan\beta$, one
gets $m^2_{H^+} \approx m_0^2 + 3 M_{1/2}^2$, while it is somewhat
reduced for large $\tan\beta$ due to negative contribution from bottom
and tau Yukawa couplings. The squark masses of the first two
generations are well approximated as $m_{\tilde Q}^2 \approx m_0^2 + 6
M_{1/2}^2$. For the third generation the expressions are somewhat
involved due to large mixing in the squared mass matrix, with the
result that one eigenvalue is lighter than any of the first two
generation squark masses. The charged Higgs loop contribution to $R_b$
is numerically sub-dominant compared to the stop loop. The limit from
$R_b$ or $A_{\rm FB}^b$, i.e. from $X_b$, effectively applies on the
stop mass, which is then translated to the charged Higgs mass. For
NUSM, the situation is different as the masses of the charged Higgs
and the stop squarks stem from different parameters, $m_{10}$ and
$m_{16}$, respectively. Still, an upper limit on the charged Higgs
mass arises, though it is much weaker than in cMSSM.  The reason is
that in NUSM the charged Higgs mass and the squark masses are not {\em
  completely} independent. As mentioned earlier, though $m^2_{H_u}$
and $m^2_{H_d}$ originate from the same $m_{10}$ at high scale, their
splitting at the weak scale following renormalization group running
picks up the squark mass dependence (essentially of the third
generation).  So, beyond a certain charged Higgs mass, the stop
squarks become too heavy to leave any impact on $X_b$. This is how an
upper limit is placed on the charged Higgs mass in NUSM, though it is
weaker than in cMSSM.

We now make some remarks on the impact of the observables $R(D)$ and
$R(D^\ast)$, defined as $R(D^{(\ast)}) = {\rm Br}\, (B\to
D^{(\ast)}\tau\nu)/{\rm Br}\, (B\to D^{(\ast)}\ell\nu)$, with
$\ell=e,\mu$. These ratios have recently been measured by the BaBar
Collaboration \cite{babar-rd}:
\begin{equation}
R(D^{(*)}) = \frac{{\rm Br} (B \to D^{(*)} \tau \nu)}{{\rm Br} (B \to
D^{(*)} \ell \nu)} = 0.440 \pm 0.058 \pm 0.042 ~(0.332 \pm 0.024 \pm
0.018)\,,
\end{equation}
which are $2.0\sigma$ and $2.7\sigma$ away from their SM estimates,
respectively.  Like $B^+\to\tau^+\nu$, the supersymmetric
contributions to these processes are completely dominated by
tree-level charged Higgs exchange. As has been observed in
Ref.~\cite{babar-rd}, consistency with $R(D)$ and $R(D^*)$ requires
$\tan\beta/m_{H^+}$ to be $(0.44 \pm 0.02)~{\rm GeV}^{-1}$ and $(0.75
\pm 0.04)~{\rm GeV}^{-1}$, respectively, ruling out any {\em otherwise
  allowed} value of $\tan\beta/m_{H^+}$ at 99.8\% confidence level.
This conclusion would hold not only in cMSSM or NUSM but in any
supersymmetric model having a single charged Higgs.

A comment on constraints from dark matter, which we have not included
in our numerical analysis, is now in order. Corners in cMSSM and NUHM
parameter spaces which contain points that satisfy both $m_h \approx$
125 GeV and the dark matter constraints have been explored, and one
such benchmark point for cMSSM parameters which satisfies LHC 2012
plus XENON100 dark matter data is: $m_0 = 389.51,~ M_{1/2} = 853.03,~
A_0 = -2664.79$ (all in GeV), and $\tan\beta = 14.50$
\cite{Strege:2012bt}. However, neither this point nor any other
benchmark point can satisfy the $X_b$ criteria.

\subsection{pMSSM}
It is perhaps appropriate at this stage to perform a statistical
analysis to convey the essence of our study.  For this, we consider
the pMSSM version of supersymmetry and exhibit how the well-motivated
5 benchmark points, shortlisted in \cite{cahill}, respond to $R_b$,
$A^b_{\rm FB}$, $R(D)$ and $R(D^*)$.  The 5 models, referenced with
identification numbers \texttt{401479}, \texttt{1046838},
\texttt{2342344}, \texttt{2387564}, \texttt{2750334}, produce ($i$)
bino-stop coannihilation and almost invisibility of the stop, ($ii$) a
pure higgsino as the lightest supersymmetric particle, ($iii$) a
compressed spectrum of squarks coannihilating with bino, ($iv$) the
$A$-funnel region with 1 TeV bino and 1.4 TeV squarks, and ($v$)
well-tempered neutralino, respectively. They satisfy all the other
experimental constraints, including the Higgs mass (all these points
produce a large stop mixing), the oblique electroweak $S$ and $T$
parameters, dark matter relic density and detection cross section.

\begin{table}[htbp]
\begin{center}
\begin{tabular}{||c|c|c|c||}
\hline
Benchmark & $\chi^2$/d.o.f & $\chi^2$/d.o.f & $\chi^2$/d.o.f\\
point & (Canonical $=$ \{C\}) & (\{C\}, $R_b$, $A^b_{\rm FB}$) & 
(\{C\}, $R_b$, $A^b_{\rm FB}$, $R(D)$, $R(D^*)$)\\
\hline
401479 &  3.73     &  3.76    &  4.21   \\
       & (0.76) & (1.99) & (3.01) \\
1046838 &  3.33    &  3.50    &  4.01   \\
       &  (0.48)  & (1.82)  & (2.88) \\
2342344 &  3.67    &  3.73    &  4.18    \\
       &  (0.51)  & (1.85) & (2.90)\\
2387564 &  3.61    &  3.68    &  4.14    \\
      & (0.69) & (1.95) & (2.97)\\
2750334 &  4.05    &  3.99    &  4.41    \\
      & (1.41) & (2.40) & (3.34) \\
\hline  
\end{tabular}
\caption{\label{tab:chisq} \small The pMSSM fit results for 5
  specified benchmark points, which predict the Higgs mass and the
  dark matter relic density in the experimentally allowed range and
  conform to electroweak precision tests. The canonical observables
  are ${\rm Br}(B_s \to \mu^+\mu^-)$, ${\rm Br}(B^+ \to \tau+\nu)$,
  ${\rm Br}(b \to s \gamma)$, and $(g-2)$ of muon.  The entries in
  parentheses correspond to fits without the $(g-2)$ of muon. For
  details, see text after Eq.~(\ref{pull_numbers}).}
\end{center}
\end{table}

Now we perform a statistical analysis with the above benchmark points
taking into account a limited set of observables for illustrative
discussion.  For
the five specified benchmark points, considered to be 5 different
models, we obtain the pulls for $R_b$, $A^b_{\rm FB}$, $R(D)$ and
$R(D^*)$ roughly as
\begin{equation}
R_b : -1.20\,, ~ A^b_{\rm FB} : 2.50\,, ~ R(D) : -1.99\,, ~ R(D^*) :
-2.67 \,.
\label{pull_numbers}
\end{equation}
We have in fact calculated these pulls for all the 24 benchmark points
\cite{cahill}, and observed that they remain the same up to two
decimal places.  Note that the changes from the SM pulls are very
marginal, and thus we can safely say that as far as these variables
are concerned, no pMSSM benchmark point performs any better than the
SM.  Next, we call a set of observables `canonical' which comprises of
${\rm Br}(B_s \to \mu^+\mu^-)$, ${\rm Br}(B^+ \to \tau+\nu)$, ${\rm
  Br}(b \to s \gamma)$, and $(g-2)$ of muon.  We first calculate the
goodness-of-fit, measured by $\chi^2$ per degree of freedom (d.o.f),
for each of the five benchmark points taking into account the
observables of the canonical set.  To gain insight into how addition
of new observables influences the fit, we increase their numbers in
two steps. First, we add $R_b$ and $A^b_{\rm FB}$ to the canonical
set, and calculate $\chi^2/$(d.o.f) for each benchmark point. Then, we
add $R(D)$ and $R(D^*)$ on top of $R_b$ and $A^b_{\rm FB}$, and check
what happens to the $\chi^2/$(d.o.f).  Each such fit is done {\em
  with} and {\em without} the $(g-2)$ of muon which is a part of the
canonical set. The fit results are displayed in Table
\ref{tab:chisq}. We highlight the salient features that come out of
this illustration.
\begin{enumerate} 
\item With the canonical set of observables, $\chi^2/$(d.o.f) les
  between 3.3 to 4.0. The fit gets even worse as we add the four new
  observables. This feature holds not only for the 5 specified
  benchmark points, but also for the larger set of 24 such points.

\item The fit improves considerably if one considers only the
  canonical set excluding the muon $(g-2)$. However, when we include
  the four new observables, exclusion of $(g-2)$ is not much of a help
  in improving the goodness-of-fit. 
\end{enumerate}
The reason behind the above behavior of the fit is not difficult to
understand. All these benchmark scenarios correspond to heavy
superparticles whose effects in virtual states are too tiny to leave
significant numerical impacts on $R_b$, $A^b_{\rm FB}$, $R(D)$,
$R(D^*)$ and the muon $(g-2)$. To sum up, supersymmetry is not any
better than the SM in resolving the tensions in the aforementioned
observables.

\section{Conclusions}
We now conclude with the following observations. If the 125 GeV scalar
resonance discovered at the LHC be after all the lightest CP-even
Higgs boson of minimal supersymmetry, in particular of the cMSSM or
NUSM variety, then accommodating all the {\em three types} of
observables simultaneously becomes extremely difficult. As mentioned
earlier, these three types are: ($i$) $m_h = (123-127)$ GeV, ($ii$)
$B$-physics observables together with the muon $(g-2)$, ($iii$) $R_b$
and $A_{\rm FB}^b$, with their pulls combined to form $X_b$.  Effects
of ($iii$) constitute the punch line of our paper.  The following
tensions are to be specially noticed in the context of cMSSM and NUSM:
($a$) Better consistency with $X_b$, which accounts for the combined
pull of $R_b$ and $A_{\rm FB}^b$, would require the stop squarks to be
light, whereas the satisfaction of the Higgs mass requires the stop
squarks to be heavy. The tension in NUSM is slightly less compared to
cMSSM. ($b$) The incompatibility between $R_b$ and $A_{\rm FB}^b$ is a
generic feature of a large class of supersymmetric models. In the
latter part of this paper, we have considered the 19-parameter pMSSM
scenario and performed a $\chi^2$ analysis with 5 experimentally
well-motivated benchmark models, which reproduce the observed Higgs
mass, the dark matter relic density and precision electroweak
observables, all consistent with experiments. We find that inclusions
of $X_b$, $R(D)$ and $R(D^*)$ adversely affect the fit. We point out
that models with very light sbottom squarks modify $R_b$ in the right
direction \cite{Arbey:2013aba}, but only at the expense of growing
disagreement with $A_{\rm FB}^b$. As far as these quantities are
concerned, one must look beyond the conventional supersymmetric
spectrum to find a compromise solution.

\noindent {\bf Acknowledgments:}~ AK was supported by CSIR, Government
of India (project no.\ 03(1135)/9/EMR-II), and also by the DRS
programme of the UGC, Government of India. The research of TSR is
supported by the Australian Research Council.

\end{document}